\documentclass[final]{agujournal2019}
\usepackage{url} 
\usepackage{lineno}

\usepackage{soul}

\usepackage[italic]{hepnames}
\usepackage{rotating}
\usepackage{tabularx,booktabs}
\usepackage[colorlinks,citecolor=blue,urlcolor=blue,linkcolor=cyan]{hyperref}
\usepackage{natbib}

\draftfalse

\journalname{Geochemistry, Geophysics, Geosystems}

\begin{document}

\title{Radiogenic power and geoneutrino luminosity of the Earth and other terrestrial bodies through time}

\authors{W. F.~McDonough\affil{1,2}, O.~\v Sr\'amek\affil{3}, and S. A.~Wipperfurth\affil{1}}

\affiliation{1}{Department of Geology, University of Maryland, College Park, MD 20742, USA}
\affiliation{2}{Department of Earth Sciences and Research Center for Neutrino Science, Tohoku University, Sendai 980-8578, Japan}
\affiliation{3}{Department of Geophysics, Faculty of Mathematics and Physics, Charles University, Prague, Czech Republic}

\correspondingauthor{W. F.~McDonough}{mcdonoug@umd.edu}

\begin{keypoints}
\item Radiogenic heat production and geoneutrino luminosity are calculated over the age of the Earth
\item Simple formulae proposed for evaluation at arbitrary planetary composition are provided
\item Compilation of both long-lived and short-lived radionuclides are provided
\end{keypoints}

\begin{abstract}
We report the Earth's rate of radiogenic heat production and (anti)neutrino luminosity from geologically relevant short-lived radionuclides (SLR) and long-lived radionuclides (LLR) using decay constants from the geological community, updated nuclear physics parameters, and calculations of the $\beta$ spectra. We track the time evolution of the radiogenic power and luminosity of the Earth over the last 4.57 billion years, assuming an absolute abundance for the refractory elements in the silicate Earth and key volatile/refractory element ratios (e.g., Fe/Al, K/U, and Rb/Sr) to set the abundance levels for the moderately volatile elements. The relevant decays for the present-day heat production in the Earth ($19.9\pm3.0$ TW) are from $^{40}$K, $^{87}$Rb, $^{147}$Sm, $^{232}$Th, $^{235}$U, and $^{238}$U. Given element concentrations in kg-element/kg-rock and density $\rho$ in kg/m$^3$, a simplified equation to calculate the present day heat production in a rock is: 
\begin{equation*}
    h \, [\mu \text{W\,m}^{-3}] = \rho \left( 3.387 \times 10^{-3}\,\text{K} + 0.01139 \,\text{Rb} + 0.04595\,\text{Sm} + 26.18\,\text{Th} + 98.29\,\text{U} \right)
\end{equation*}
The radiogenic heating rate of Earth-like material at Solar System formation was some 10$^3$ to 10$^4$ times greater than present-day values, largely due to decay of $^{26}$Al in the silicate fraction, which was the dominant radiogenic heat source for the first $\sim10$\,Ma. Assuming instantaneous Earth formation, the upper bound on radiogenic energy supplied by the most powerful short-lived radionuclide $^{26}$Al (\textit{t$_{1/2}$} = 0.7 Ma) is 5.5$\;\times\;$10$^{31}$ J, which is comparable (within a factor of a few) to the planet's gravitational binding energy.  
\end{abstract}

\section*{Plain Language Summary}
The decay of radioactive elements in planetary interiors produces heat that drives the dynamic processes of convection (core and mantle), melting and volcanism in rocky bodies in the solar system and beyond. For elements with half-lives of 100,000 to 100 billion years, uncertainties in their decay constants range from 0.2$\%$ to $\sim4\%$, and comparing data from physics versus geology show differences about 1$\%$ to 4$\%$. These differences, combined with uncertainties in $Q$ values (energy released in reaction), lead to diverging results for heat production and for predictions of the amount of energy removed from the rocky body by emitted (anti)neutrinos.

\section{Introduction}

Radioactive decay inside the Earth produces heat, which in turn contributes power to driving the Earth's dynamic processes (i.e., mantle convection, volcanism, plate tectonics, and potentially the geodynamo). The physics community, using the latest numbers from nuclear physics databases, provides estimates of the present-day radiogenic heating rate and geoneutrino luminosity (i.e., number of particles per unit of time) of the Earth \citep{Enomoto2006, Fiorentini2007b, Dye2012, Usman2015, Ruedas2017}. These studies include comprehensive reviews of the fundamental physics of these decay schemes, covering both the energy added to the Earth and that removed by the emitted geoneutrinos. This note draws attention to differences in decay constants ($\lambda$) as reported in the geological and physics literature and recommends the former as being more accurate and precise. In doing so, we recognize that the latter sources critically combine data from the physics, geology, and cosmochemical communities. Recent papers \citep[e.g.,][]{Villa2015,Villa2016,Naumenko2018} provide examples of where geochronologists have cross calibrated multiple decay chains and provide details of their meta-analysis of data and independent reassessments of uncertainties. The absolute accuracy of geological studies is underpinned by the $^{238}$U decay constant \citep{Jaffey1971}, which is a single measurement from a 50 year old study. More recently, \cite{schoene2006reassessing,Villa2016,parsons2018new} are working to refine $\lambda$ values for {$^{238}$U}, {$^{235}$U}, and {$^{234}$U} with reduced experimental uncertainties. The relative accuracies for various $\lambda$ values are based on multiple cross-calibrations for different decay systems on the same rocks and mineral suites. Improvements in measurement precision come from repeated chronological experiments.

There are a number of naturally occurring short-lived (relative to the Earth's age; half-lives $t_{1/2}<10^8$\,years) and long-lived ($t_{1/2}>10^9$\,years) radionuclides; those discussed here have half-lives between $10^5$ and $10^{11}$ years. The long-lived decay constants are listed in Table~\ref{tab:systems} along with their decay modes and decay energies. The decay modes include alpha ($\alpha$), beta-minus ($\beta^-$), and electron capture (\textrm{EC}). The beta-plus ($\beta^+$) decay mode is less common, but is seen in the $^{26}$Al system, as well as a few minor branches in the Th and U decay chains, and also likely in the $^{40}$K branched decay. Geoneutrinos are naturally occurring electron antineutrinos ($\APnue$) produced during $\beta^-$ decay and electron neutrinos ($\Pnue$) produced during $\varepsilon$ (i.e., $\beta^+$ and EC) decays. We recognize that the use of $\varepsilon$ to refer the combination of EC and $\beta^+$ is not uniformly accepted. However, we are following the recommendation of the NuDat2 glossary 
(\url{https://www.nndc.bnl.gov/nudat2/help/glossary.jsp#nucleardecay}). The generic versions of these decay schemes are:

\begin{equation} 
\begin{split} 
\textrm{Alpha} \quad (\alpha)\qquad   &^A_ZX \rightarrow \;^{A-4}_{Z-2}X' + \alpha +Q, \\
\textrm{Beta Minus} \quad (\beta^-) \qquad &^A_ZX \rightarrow^A_{Z+1}X' + e^- + \APnue + Q, \\
\textrm{Electron Capture} \quad (\textrm{EC}) \qquad  &^A_ZX + e^- \rightarrow ^A_{Z-1}X' + \Pnue + Q, \\
\textrm{Beta Plus} \quad (\beta^+) \qquad &^A_ZX \rightarrow^A_{Z-1}X' + e^+ + \Pnue + Q, \\
\varepsilon \qquad & \textrm{refers to the combination of EC and } \beta^+
\end{split}
\end{equation}

\noindent with parent element $X$, daughter element $X'$, mass number $A$, atomic number $Z$, energy of reaction $Q$, electron $e^-$, positron $e^+$, and alpha particle $\alpha$ ($^{4}_{2}$He nucleus).


We report radiogenic heat production and (anti)neutrino luminosity from geologically relevant short-lived radionuclides (SLR) and long-lived radionuclides (LLR). For the LLR we compare half-lives used in the geological and nuclear physics communities and recommend use of the former. We calculate the heat added to the Earth by these nuclear decays, as well as the energy carried away by (anti)neutrinos that leave the Earth. We calculate estimates of the embedded and removed energy of decay, particularly for the SLR, from $\beta$ decay spectra calculated using Fermi theory and shape factor corrections. We conclude by presenting models for the Earth's radiogenic power and geoneutrino luminosity for the last 4567 million years, along with simple rules for extrapolating these results to other terrestrial bodies and exoplanets.

\section{Contrasting methodologies}

In compiling the data needed to calculate all of the observables, we found differences between the decay constants ($\lambda=\frac{\ln 2}{t_{1/2}}$) reported by the geological and nuclear physics communities.  Values for extant systems are provided in a side-by side comparison  in Table~\ref{tab:comparison}. The rightmost column reports the relative difference, in percent, between the decay constants from these communities and for some, the difference can be considerable (more than 30\%). An updated physics number for the half-life of $^{190}$Pt reported in \cite{Braun2017} agrees with the numbers obtained by \cite{Cook2004}, who presented a detailed study of a suite of well behaved (closed system evolution), 4.5 billion year old, iron meteorites (i.e., group IIAB and IIIAB).

There is a 1.1\% difference in the decay constant for $^{40}$K between literature sources, which is a nuclide that provides $\sim\!20\,\%$ of the planet's present-day radiogenic heat and $\sim\!70\,\%$ of its geoneutrino luminosity (see Table~\ref{tab:hprod}). This difference is outside of the uncertainty limits on the half-life of $^{40}$K, recently established by geochronologists \citep{Renne2011}. Differences in decay constants directly scale (i.e., 1.1\%) into increases or decreases in radiogenic power and geoneutrino luminosity.


Differences in decay constants reported by the geological and nuclear physics communities come from the methods used to establish the absolute and relative half-lives. Physics experiments typically determine a half-life value by measuring the activity $A = -dN/dt = \lambda N$ ($N$ is the number of atoms) of a nuclide over time, whereas geochronology studies empirically compare multiple decay systems for a rock or suite of rocks that demonstrate closed system behavior (show no evidence of loss of parent or daughter nuclide). The number of atoms $N$ of parent nuclide evolves according to $N=N_0 e^{-\lambda t}$, therefore $\ln N = \ln N_{0} - \lambda t$. A plot of $\ln N$ (ordinate) vs. $t$ (abscissa) gives a line of slope $-\lambda$ with y-intercept equal to $\ln N_0$.

Direct counting experiments generally involve the isolation of a pure mass of the parent nuclide of interest, knowing exactly the number of parent atoms at the start of the experiment, and then determining the ingrowth of daughter atoms produced at one or more times later \citep{Begemann2001}. Measuring the decay rate can be accomplished by either detecting emitted energetic particles (i.e., $\alpha$, $\beta$, or $\gamma$) or directly measuring the amount of daughter atoms produced after a considerable period of time (years to decades). The latter method is particularly useful for low energy $\beta$ emitters (e.g., $^{87}$Rb and $^{187}$Re).

Geochronological experiments compare multiple chronometric methods (e.g., U--Pb and K--Ar systems \citep{Renne2011,Naumenko2018}) and develop a series of cross calibrations, where the shortcoming of this approach is the anchoring decay system that pins down the accuracy for other chronometers. Table~\ref{tab:comparison} highlights the differences in half-life values reported in a standard physics reference source NNDC (National Nuclear Data Center, \url{http://www.nndc.bnl.gov}) and geology. Relative differences at the $\sim1\%$ scale and greater are seen for $^{40}$K, $^{87}$Rb, $^{176}$Lu, $^{187}$Re and $^{190}$Pt decay systems.

Radioactive decay involves the transition to a lower level energy state of a nuclear shell and the accompanying release of energy, requiring the conservation of energy, linear and angular momenta, charge, and nucleon number. The kinetic energies of emitted alpha particles are discrete and on the order of 4 to 8 MeV, whereas different forms of beta decay show a continuous spectrum with characteristic mean and maximum energies for a given decay and the (anti)neutrino carrying away a complementary part of the energy. The energy of the beta decay process is partitioned between the electron, the antineutrino (or positron and neutrino), and the recoiling nucleus. Differences in heat production per decay reported in different studies are largely due to differences in decay energies (minimal differences), the energy carried off by (anti)neutrinos, and the branching fractions in the case of branched decays (large differences for the latter two). Furthermore, the time rate of heat production is sensitive to the value of the decay constant. This study differs from other recent efforts \citep{Dye2012, Ruedas2017, Usman2015, Enomoto2006, Fiorentini2007b} in its input assumptions; we use decay constants and branching fractions from geochronological studies and we calculate the beta decay energy spectrum for most of the SLR and $^{40}$K decays. For the remaining LLR decays, we adopt the energy spectra from \cite{enomoto:nuspec}. 

\begin{figure}[h]
\centering
\includegraphics[scale=1]{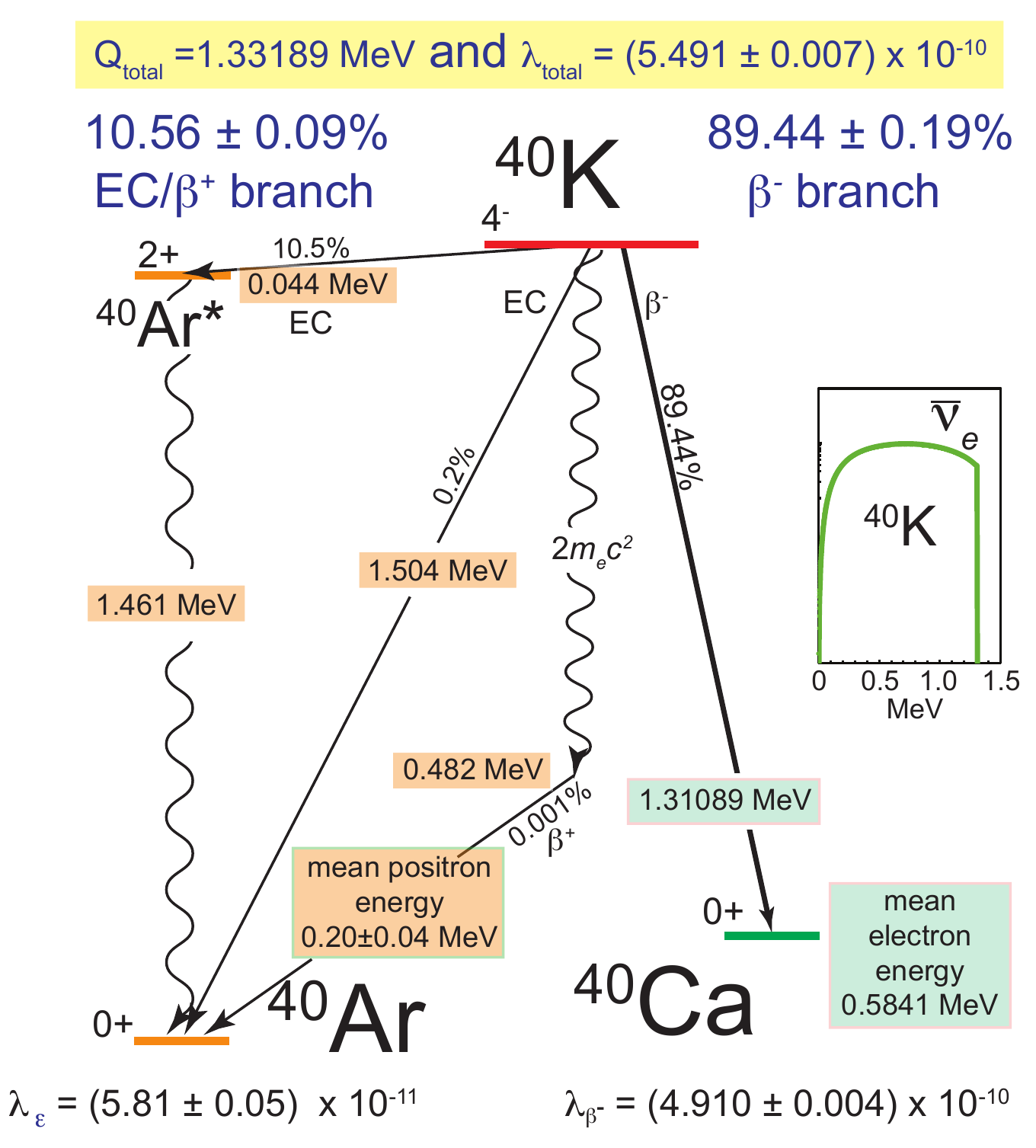}
\caption{Decay scheme for $^{40}$K. The beta-minus branch directly leads to $^{40}$Ca in the ground state accompanied by the emission of an $\APnue$, whereas the electron capture branch has the emission of a 44 keV $\Pnue$ to the excited state of $^{40}$Ar*, with the latter undergoing an isomeric transition to the ground state of $^{40}$Ar via the emission of a 1.46 MeV $\gamma$-ray. Minor branches that we account for include the electron capture and $\beta^+$ to ground state of $^{40}$Ar. During $\beta^-$ decay the energy is shared between the $\Pelectron$ and $\APnue$, with the latter particle removing on average 650\,keV of energy from the Earth (accounting for branching; the mean $\APnue$ energy is 727\,keV). The $\beta^+$ to ground state transition is noted by 2m$_e$c$^2$, accounting for the $\gamma$ photons from $\Pelectron$ -- $\APelectron$ annihilation, and energy shared between the $\APelectron$ and $\Pnue$.  Branching ratios and uncertainties are from \cite{Naumenko2018}, with additional insights from \cite{Renne2011}. Half-lifes and $^{40}$K abundance are from \cite{Naumenko2013}. Decay energies are either calculated (e.g., $\beta^-$ from antineutrino spectrum) or from \cite{Chen2017}, with the latter reference as the source for angular momentum and spin parity states. The antineutrino energy spectrum (with intensity in arbitrary units) shown in the inset uses the $\beta^-$ shape factor from \cite{Leutz1965} to account for the correction of the third unique forbidden transition.}
\label{fig:k40}
\end{figure}

The $^{40}$K decay scheme is a good example of where differences in inputs occur. Many naturally occurring decay schemes have a single decay mode, whereas $^{40}$K is a branch decay scheme with $\beta^-$ and $\varepsilon$ decays (see Figure~\ref{fig:k40}), with emission of an $\APnue$ and $\Pnue$, respectively, removing energy from the Earth. The amount of radioactive heating in the Earth from this branch decay scheme depends on the branching ratio and the energy carried by the $\Pnue$ and $\APnue$.
Given their initial criteria of coincident Rb-Sr and U-Pb ages for minerals from the Phalaborwa carbonatite complex and assuming a range of branching ratios on the basis of published literature values, \cite{Naumenko2018}  calculated a $\lambda_{Total}$ for $^{40}$K based on their K-Ca isochron and reported a probability for the $\beta^-$ branching between 89.25\,\% and 89.62\,\% and for the $\varepsilon$ branching between 10.38\,\% and 10.75\,\%. They highlighted that the errors on these branching ratio values are non-Gaussian. The physics community reports the branching probabilities of $\beta^-$ as 89.28(11)\,\% and of $\varepsilon$ as 10.72(11)\,\% \citep{Chen2017}. Figure~\ref{fig:k40} reports the updated $^{40}$K decay scheme---the branching fractions, the average energies removed by the antineutrinos and neutrinos, and the energy deposited by these decays.


Beta decay involves transforming a quark state in the nucleus and emission of a pair of fermions ($e^- \APnue$ or  $e^+ \Pnue$), where each fermion has an intrinsic angular momentum (or ``spin'') of 1/2. The decay satisfies all the relevant conservation principles of particle physics, including the electron-lepton number ($L_e$) conservation, where $L_e=1$ for matter particles ($e^-,\Pnue$) and $L_e=-1$ for antimatter particles ($e^+,\APnue$). The transformation is accompanied by a change in the total angular momentum of the nucleus ($\Delta I$) which, by conservation of angular momentum, must be reflected in the state of the $e\nu$ pair, that is, the total orbital angular momentum ($L$) and the total spin angular momentum ($S^L$) of the $e\nu$ pair. 

Beta decays can be pure Fermi transitions, pure Gamow-Teller transitions, or a combination of both. In Fermi transitions the spins of the emitted leptons are anti-parallel, $S^L=0$, and therefore $\Delta I$ is matched only by $\pm L$. In Gamow-Teller transitions the spins of the emitted leptons are aligned, i.e., $S^L=1$, and coupled to the change in nuclear angular momentum state $\Delta I$ together with $L$: $\Delta I=\pm|L\pm1|$. The so-called ``unique'' transitions are Gamow-Teller transitions where $L$ and $S^L$ are aligned, $\Delta I=\pm|L+1|$. Typically, transitions with a higher $L$ have a longer half-life ($t_{1/2}$), because of less overlap of the $e\nu$ wave functions with the nucleus. Transitions with a non-zero $L$ are called ``forbidden'' (as opposed to ``allowed'' for $L=0$), which really means suppressed decays that involve changes in nuclear spin state; in $n$-th forbidden transitions the $e\nu$ pair carries $n$ units of orbital angular momentum \citep{bielajew:2014notes}. For example, the $^{40}$K decay scheme involves a third unique forbidden transition, whereas the $^{87}$Rb decay scheme involves a third non-unique forbidden transition.

Following Fermi's theory and working in units $\hbar=m_e=c=1$, the shape of a $\beta$ spectrum is calculated from

\begin{equation} \label{spectrum}
\frac{dN}{dw} \propto pwq^2 F(Z,w) S(w)
\end{equation}

\noindent and normalized to the branching fraction of the specific $\beta$ decay \citep{enomoto:phd}. In equation \eqref{spectrum} $w=1+E$ is the total energy of the $\beta$-particle ($E$ being its kinetic energy), $p=\sqrt{w^2-1}$ is the momentum of the $\beta$-particle, $q$ is the total energy of the neutrino (equal to its momentum as the neutrino mass is negligible) satisfying $E+q=E_\text{end}$, where $E_\text{end}$ is the endpoint energy of the transition (in the case of a transition to ground state, it is the $Q$-value), and $Z$ is the charge of the daughter nucleus. The left-hand side of equation \eqref{spectrum} is the probability of a $\beta$ particle to be created with energy in the $dw$ vicinity of $w$, where $w$ goes from 1 to $1+E_\text{end}$. The right-hand side is a product of three factors, the phase space factor $pwq^2$, the Fermi function $F(Z,w)$, and the shape factor $S(w)$. The Fermi function

\begin{equation} \label{fermi}
F(Z,w) \propto (w^2-1)^{\gamma-1} \mathrm{e}^{\pi\eta} \big|\Gamma(\gamma+i\eta)\big|^2,
\end{equation}

\noindent where

\begin{eqnarray}
\gamma & = & \sqrt{1-(\alpha Z)^2},\\
\eta & = & \frac{\alpha Z w}{\sqrt{w^2-1}},
\end{eqnarray}

\noindent $\alpha$ being the fine-structure constant, accounts for the Coulombic interaction between the daughter nucleus and the outgoing $\beta$-particle and the gamma function $\Gamma$ takes a complex argument \citep{enomoto:phd}. The shape factor $S(w)$, often written at $S(p,q)$, is equal to 1 for allowed transitions and has a more complex energy-dependence in the case of forbidden transitions.

A review of many $\beta^-$ decay energy spectra was recently given by \cite{Mougeot2015}, including the shape factors used for the forbidden transitions. We adopt these shape factors in our calculations, but also include additional $\beta$ decays not studied by \cite{Mougeot2015}; the shape factors used here are listed in Table~\ref{tab:ShortLived}. We have performed the $\beta$ spectra evaluation and calculated the average energy removed by the $\Pnue$ and $\APnue$, which are reported as $Q_\nu$ (MeV) in Table~\ref{tab:hprod} and can be calculated from $Q-Q_h$ in Table~\ref{tab:ShortLived}.

\section{Radiogenic heat and geoneutrino luminosity of the Earth}

Using decay constants for short-lived and long-lived radionuclides and $^{40}$K branching ratio from the geological literature we calculate the heat production and geoneutrino luminosity of the bulk silicate Earth (BSE) based on a model composition (Tables~\ref{tab:hprod} and \ref{tab:ShortLived} and references therein). Compositional models differ on the absolute amount of refractory elements (e.g., Ca and Al) in the Earth \citep[see reviews in][]{McDonough2014,McDonough2016}, which includes La, Sm, Lu, Re, Pt, Th, and U. The model composition for the BSE fixes the absolute abundances of the refractory elements at 2.75 times that in CI1 chondrites \citep{McDonough1995}. Compositional models for the bulk Earth and core are from \cite{McDonough2014}. Radiogenic elements in the core include the short-lived (e.g., $^{60}$Fe, $^{79}$Se, $^{97,98,99}$Tc, $^{106}$Pd, and to a lesser extent $^{53}$Mn, ($\pm\;^{92}$Nb, which might or might have been partitioned into the core), $^{126}$Sn and $^{205}$Pb) and long-lived isotopes (i.e., $^{187}$Re, and $^{190}$Pt), whereas other, nominally lithophile elements, with negligible siderophile tendencies, are considered to have insignificant core concentration levels for this calculation. For critical volatile elements, there is a reasonable consensus to use their ratios with refractory elements. For example, \cite{Arevalo2009} reported the K/U value for the silicate Earth as $13,800\pm1,300$ (1 standard deviation). Constraints for Rb come from the constancy of the Ba/Rb and the Sr--Nd isotopic system \citep[assumes the BSE has an $^{87}$Sr/$^{86}$Sr between 0.7040 and 0.7060, based on the mantle array;][]{Hofmann2014} and the Rb/Sr values (Ba and Sr are refractory elements with abundances set at 2.75 times that in CI1 chondrites) for the bulk silicate Earth, leading to a Rb/Sr of $0.032\pm0.007$ \citep{McDonough1992c}.

\begin{figure}[h]
\centering
\includegraphics[width=\textwidth]{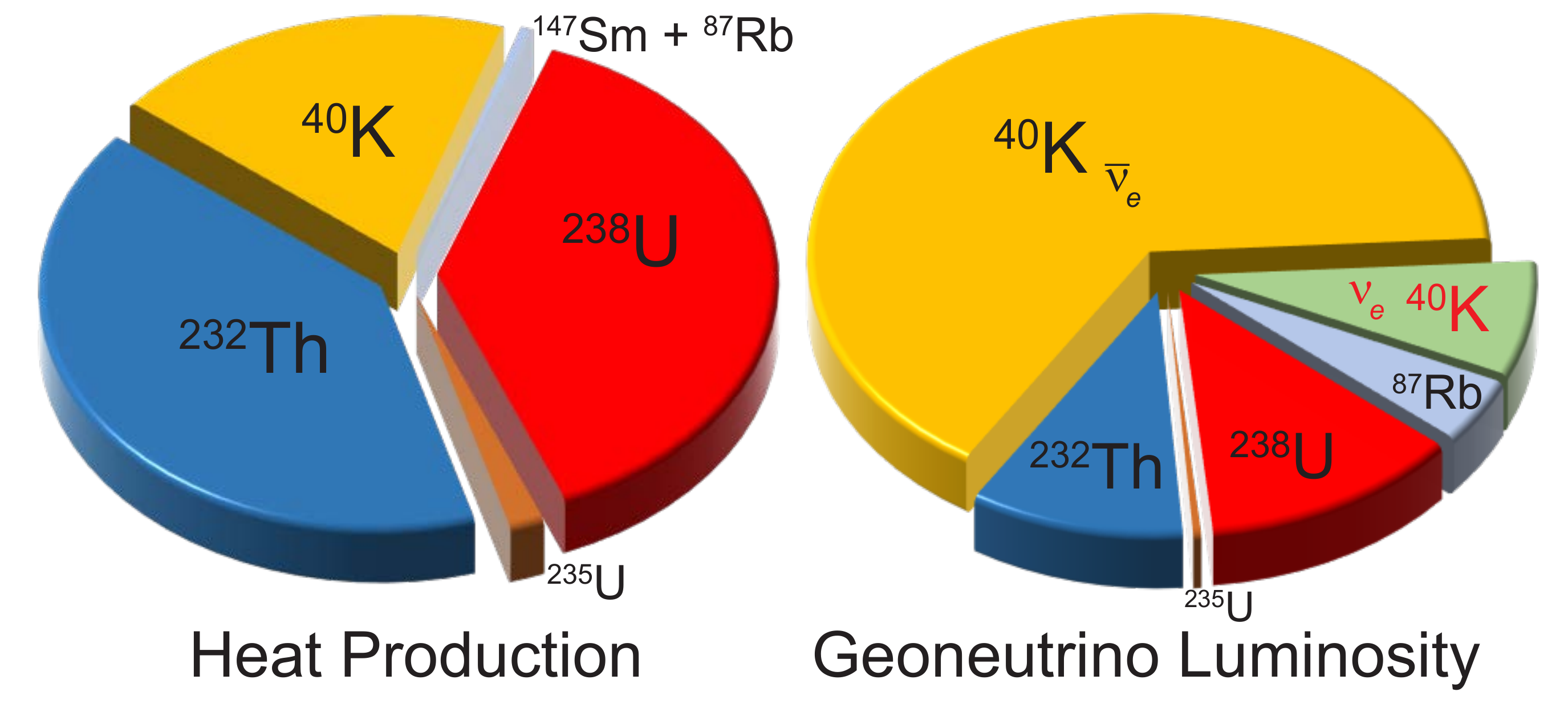}
\caption{The relative contributions to radiogenic heat production and geoneutrino luminosity of the present-day Earth. Note the relative contributions of $\APnue$ and $\Pnue$ from $^{40}K$ in terms of geoneutrino luminosity. (Antineutrinos emitted by human-made nuclear reactors are not considered.)}
\label{fig:pie}
\end{figure}

Heat production and geoneutrino emission data for $^{40}$K, $^{87}$Rb, $^{147}$Sm, $^{232}$Th, $^{235}$U, and $^{238}$U are reported in Table~\ref{tab:hprod}, as these are the most significant present-day producers within the Earth. In fact, 99.5\,$\%$ of the Earth's radiogenic heat production comes from $^{40}$K, $^{232}$Th, $^{235}$U, and $^{238}$U alone. The fractional contributions to heat production from $^{138}$La, $^{176}$Lu, $^{187}$Re, and $^{190}$Pt add up to $<3\times10^{-5}$ of the total radiogenic heat and 1\,\% of the Earth's geoneutrino luminosity, with virtually all of this latter minor contributions coming from $^{187}$Re. Figure~\ref{fig:pie} illustrates the present day relative contributions of heat production and geoneutrino luminosity from the major radionuclides reported in Table~\ref{tab:hprod}.



A simple formula for the present-day radiogenic heating rate $\tilde h$ (in microwatts ($\mu$W) per kg of rock) or \textit{h} (in $\mu$W m$^{-3}$ of rock) from long-lived radionuclides is presented in equations \eqref{specheat} and \eqref{specheat-h}, respectively, where $A$ is elemental concentration as mass fraction (kg-element/kg-rock; e.g., K is mass fraction of potassium), and the remaining parameters combine into numerical factors whose values are set ($N_A$ is Avogadro's number (atoms/mole), $X$ is natural molar isotopic fraction, $\mu$ is molar mass of element (kg/mole), $\lambda$ is decay constant (s$^{-1}$), $Q_h$ is radiogenic heat released per decay (J), $\rho$ is rock density (kg m$^{-3}$)). Multiplying with the mass of the geochemical reservoir of interest $M_\text{res}$ (to which the elemental concentrations apply), one gets the total radiogenic power $H$ (in terawatts) in that reservoir as shown in equation \eqref{twheat}.
Similarly, the natural specific antineutrino and neutrino luminosities $\tilde l$ (in number of particles per second per kilogram of rock) are calculated from equations \eqref{speclumanu} and \eqref{speclumnu}. Multiplication with a reservoir mass gives the total luminosities $L_{\APnue}$ and $L_{\Pnue}$ (equation \ref{lumnu}; contributions from individual elements listed in Table~\ref{tab:hprod}).

\begin{eqnarray}
\tilde h~[\text{$\mu$W\,kg}^{-1}] & = & \sum\limits_\text{LLRs} \frac{N_A X \lambda Q_h}{\mu} A \label{pre-specheat}\\
 & = & \left( 3.387 \times 10^{-3}\,\text{K} + 0.01139 \,\text{Rb} + 0.04595\,\text{Sm} + 26.18\,\text{Th} + 98.29\,\text{U} \right) \label{specheat}\\
 h~[\text{$\mu$W\,m}^{-3}] & = & \rho \times \tilde h  
 \label{specheat-h}\\
H~[\text{TW}] & = & \tilde h \times M_\text{res} \times 10^{-18} \label{twheat}\\[15pt]
\tilde l_{\APnue}~[\text{s}^{-1}\,\text{kg}^{-1}] & = & \sum\limits_\text{LLRs} \frac{N_A X \lambda n_{\APnue}}{\mu} A = \left( 2.797\,\text{K} + 86.82\,\text{Rb} + 1617\,\text{Th} + 7636\,\text{U} \right) \times 10^4 \label{speclumanu}\\[5pt]
\tilde l_{\Pnue}~[\text{s}^{-1}\,\text{kg}^{-1}] & = & \sum\limits_\text{LLRs} \frac{N_A X \lambda n_{\Pnue}}{\mu} A = 0.3302\,\text{K} \times 10^4 \label{speclumnu}\\[5pt]
L~[\text{s}^{-1}] & = & \tilde l \times M_\text{res} \label{lumnu}
\end{eqnarray}

To understand the evolution of the Earth's radiogenic heat and geoneutrino luminosity we must understand the initial starting abundances of the SLR in the solar system (listed in Table~\ref{tab:ShortLived}). At 4.57 Ga the local interstellar medium was populated with gas-dust clouds that were likely in secular equilibrium with ambient galactic sources prior to solar system formation. Recent calculations by \cite{Wasserburg2017} demonstrate that the proportional inventory of $^{26}$Al, $^{60}$Fe, $^{107}$Pd, and $^{182}$Hf in the early solar system is unlikely to be a product from an asymptotic giant branch (AGB) star. Moreover, supernova sources would likely provide abundant $^{26}$Al and $^{60}$Fe, whereas the early solar system content of $^{60}$Fe is equivalent to the measly ambient galactic supply \citep{Trappitsch2018}. More recent suggestions envisage stellar winds from a massive Wolf-Rayet star injecting $^{26}$Al to complement the local inventory of ambient galactic sources \citep{Young2014,Gounelle2012,Dwarkadas2017}. At the same time, the enhanced abundance of $^{53}$Mn and the presence of very short half life isotopes (e.g.,$^{41}$Ca {\it t$_{1/2}$} = 0.1 Ma) present challenges to be explained by models invoking Wolf-Rayet stars \citep{Vescovi2018}. Thus, the addition of mass and momentum from such a stellar source could cause a gravitational collapse of a molecular gas-dust cloud, which may have triggered our solar system formation and explain the observed proportions of short-lived radionuclides.

\begin{figure}[h]
\centering
\includegraphics[width=1\textwidth]{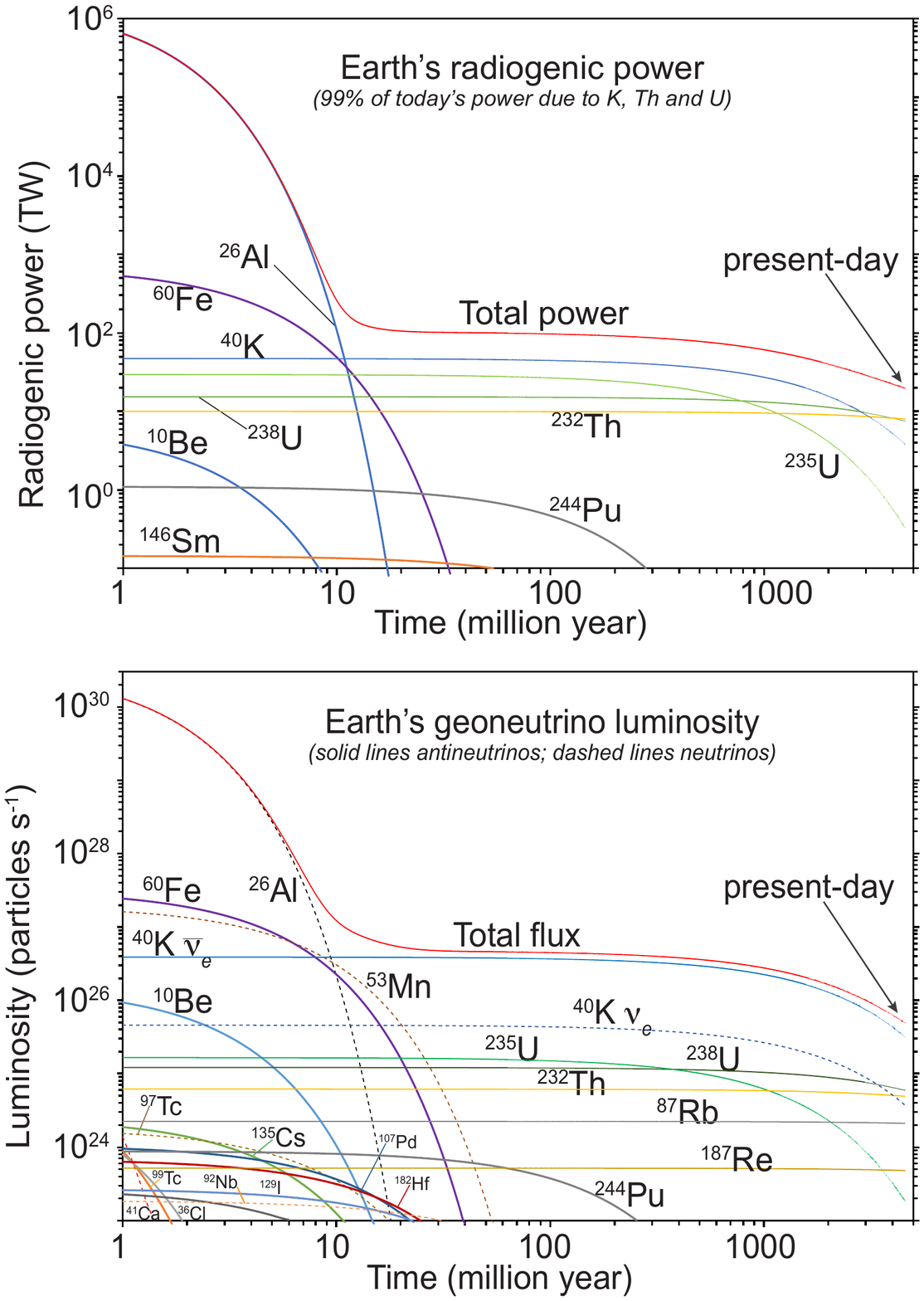}
\vspace{-15pt}
\caption{The Earth's radiogenic power (upper panel) and geoneutrino luminosity (lower panel) over the last 4567 million years. The compositional model for the bulk Earth is from \cite{McDonough2014} and concentrations are calculated back in time from present concentrations and initial ratios reported in Table~\ref{tab:ShortLived}. These figures assume an Earth mass of $6\times10^{24}$~kg at all times. The power and geoneutrino flux is scalable; if one assumes 1/10 the planetary mass, it has 1/10 the power and luminosity, for an Earth bulk composition.
}
\label{fig:flux}
\end{figure}

The total heat production and geoneutrino luminosity for models of the Earth are plotted with respect to time in Figure~\ref{fig:flux}, which were calculated using inputs from Tables~\ref{tab:hprod} and~\ref{tab:ShortLived}, expanded versions of equations (\ref{specheat}--\ref{lumnu}), including the time dependence, all the short-lived radionuclides, and updated values for the BSE and core \citep{McDonough1995,Arevalo2009,McDonough2014,McDonough2017core,wipperfurth:2018}. We assume spatial homogeneity in the protoplanetary disk for the distribution of the short-lived radionuclides (SLR) \citep[\textit{cf.},][]{larsen2011evidence,Liu2017}. This figure presents a simple illustrative example of the Earth's heat production and geoneutrino luminosity that assumes full mass at 1 million years after after solar system initiation. Of course, the Earth's heat production and geoneutrino luminosity can only be measured today, but, as can be seen in Figure~\ref{fig:flux}, these parameters have changed considerably over time.

The uncertainties for the BSE abundances reported in Table~\ref{tab:hprod} are $\pm10\,\%$ for K and the refractory lithophile elements \citep{wipperfurth:2018}, with correlations between K, Th and U. Using this Earth model and assuming negligible radiogenic heat production in the core, the present day's fluxes are $19.9\pm3.0$\,TW (terawatts or $10^{12}$\,watts) for radiogenic heat and the total geoneutrino luminosity is $(4.89\pm0.74)\times10^{25}\,(\APnue+\Pnue$)\,s$^{-1}$. The results shown in Figure~\ref{fig:flux} are directly scalable for different size planetary bodies with a bulk Earth composition; lowering the mass of a planet by a factor of 10 results in a decease by a factor of 10 in the heat production and (anti)neutrino luminosity. The most important factors are the amount of refractory elements and the volatility curve for the planet. The Earth has an Fe/Al value of $20\pm2$ \citep{McDonough1995,Allegre1995}, comparable to the chondritic ratio, which is $19\pm4$ (less the 35 value for EH chondrites). The Fe/Al value sets the proportion of refractory elements (Al) to one of the 4 major elements (i.e., O, Fe, Mg and Si) that make up $\sim\!93\,\%$ of the mass of a terrestrial planet. These latter elements are not in fixed chondritic proportions, as is the case for the refractory elements, thus, the mass proportion of O, Fe, Si and Mg can be approximated as 30:30:20:20 (or 50:15:15:15 for atomic proportions), respectively, with proportional differences leading to variations in the metal/silicate mass fraction and fraction of olivine (Mg$_2$SiO$_4$) to pyroxene (MgSiO$_3$) in the silicate shell. A K/U or K/Th value sets the volatile depletion curve for the planet. Using $\tilde h$/$A$ factors given in Table~\ref{tab:ShortLived} and expanded versions of equations~\eqref{specheat} and~\eqref{twheat}, including the time dependence, one can calculate the radiogenic power supplied to growing rocky planetary bodies of various final sizes, both within and external to the Solar System. Obviously, this cannot substitute for careful modeling of overall thermal history, which involves the additional complexities of accretion and the supplied gravitational energy, the surface boundary condition, and possible further effects of internal dynamics \citep[e.g.,][]{Sramek2012}.

We also compare our results for the present-day radiogenic power and geoneutrino flux with values reported in the literature (Table~\ref{tab:result}). Where possible, we used the abundances and masses reported in Table~\ref{tab:hprod} to carry out these comparisons. The 1.2\% spread in estimates of the BSE radiogenic power in these models \citep{Enomoto2006,Dye2012,Ruedas2017} approaches the propagated uncertainties for the decay constants, the largest source of error, and a minor change in the $^{40}$K branching ratio, which results in the emitted ${\APnue}$ removing $\sim\,$1.5\% more energy.  Marked differences are seen for the antineutrino luminosity ($L_{\APnue}$ from K, Rb, Th and U) and neutrino luminosity ($L_{\Pnue}$ from K) of the different BSE models \citep{Enomoto2006,Dye2012,Usman2015} (see Table~\ref{tab:result}). The largest discrepancy is with the \cite{Enomoto2006} study, where he reports (his Table 1) a factor of 10 lower values for the $L_{\APnue}$/(kg s) from $^{235}$U and $^{40}$K (middle column), but reports the correct total luminosity for $^{235}$U in his final column (total $L_{\APnue}$). In the \cite{Usman2015} study these authors calculated the BSE composition based on existing KamLAND and Borexino detector data at the time and lithospheric models from  \cite{huang2013reference}, and did not include a contribution from $^{87}$Rb $L_{\APnue}$.  The closest match to this study is \cite{Dye2012} where the differences are due to a 3.5\% lower prediction for $^{40}$K $L_{\APnue}$ (\textit{cf.}, molar mass and decay constant) and no contribution from $^{87}$Rb $L_{\APnue}$.


\section{Secular variation in the heat and luminosity of the Earth}

\begin{figure}[h]
\centering
\includegraphics[width=1\textwidth]{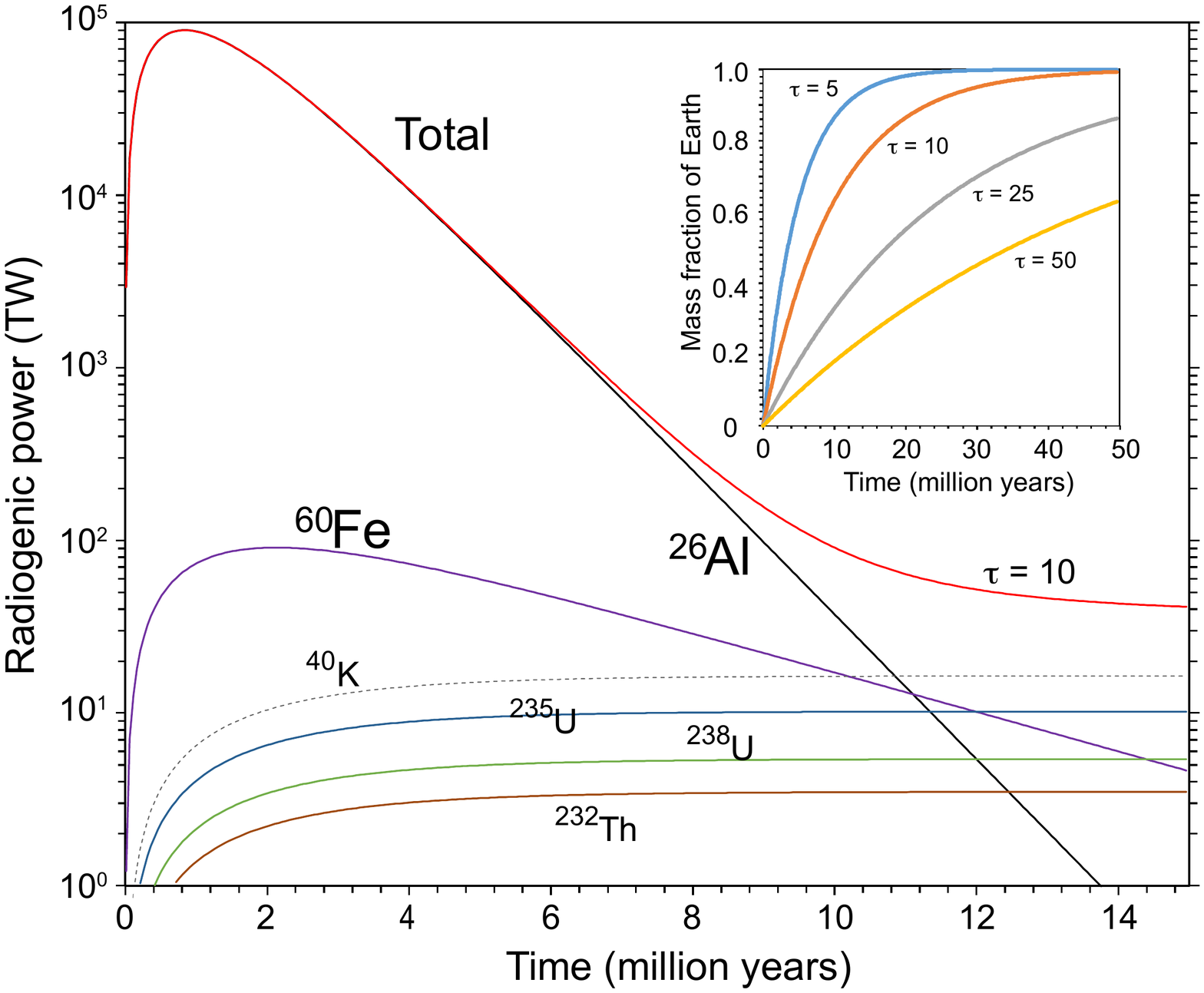}
\caption{A plot of the relative contributions of radiogenic heat to the Earth during accretion over the first 15 million years of Solar system history. The compositional model for the bulk Earth is from \cite{McDonough2014}. Figure~\ref{fig:flux} shows that other short-lived radionuclides contribute negligible amounts of power than what is shown here. Inset diagram shows a series of exponential growth curves $M(t)/M_\text{final}=1-\exp{(-t/\tau)}$ for planets. Given an age of Mars of between 2 and 5 million years \citep{Dauphas2011,Bouvier2018}, its accretion history can be modeled assuming $\tau\leq5$. For the Earth we assume $\tau=10$, however the absolute $\tau$ value is not significant, as there is only a 40\% reduction in radiogenic power at the peak between a Mars \citep{yoshizaki2020composition} and Earth accretion model.
}
\label{fig:accretion}
\end{figure}

Secular evolution of the Earth's heat production reveals that only two of the short-lived radionuclides, $^{26}$Al, and $^{60}$Fe, contribute any significant amount of additional heating to the accreting Earth above the power coming from the long-lived radionuclides (Figure~\ref{fig:accretion}). Formation and growth of the Earth is envisaged as a process that occurred on timescales of $10^7$ years. Planetary growth initiated from planetesimal "seeds" that were $10^2$ km in scale and likely formed contemporaneously with CAI (Calcium Aluminum Inclusion) formation at $t_\text{zero}=t_\text{CAI}$ (i.e., oldest known materials in the solar system) or shortly thereafter. 

The inner solar system (circa inside of 4 to 5 AU), the domain of the terrestrial planets and rocky asteroids, has been characterized as home to the NC meteorites, the non-carbonaceous meteorites \citep{warren2011stable}. Recent findings from various isotope studies of iron meteorites \citep{Kruijer2014,Kruijer2017,hilton2019} show that many of these "NC" bodies formed contemporaneously with and up to 1 Ma after $t_\text{CAI}$ formation \citep{hilton2019}. The CC iron meteorites, those relating to carbonaceous chondrites, appear to have formed slightly later (0.5 to 1 Ma) and further out, beyond 4 AU \citep{Kruijer2017,hilton2019}.

It is generally concluded that the larger of these early formed planetesimals rapidly grew in a runaway growth phase followed by oligarchic growth to where they reached Mars and Mercury size bodies, however, if the growth proces also included pebble accretion it can occur faster \citep{Izidoro2018,Johansen2017}. The mean timescales $\tau$ for terrestrial planet formation (corresponding to accretion of $\sim\!63\,\%$ of the planet's mass and assuming a simple parameterized version of planetary mass growth $M(t)/M_\text{final} = 1-\exp(-t/\tau)$) are not well constrained. Mars is suggested to have a $\tau$ of  $\sim\!2$ million years after $t_\text{CAI}$, coincident with core formation \citep{Dauphas2011,TANG2014}, meaning it likely formed within the lifetime of the protoplanetary disk. Formation timescales also depend on position in the disk \citep{Johansen2017}. \cite{Izidoro2018} found, depending on the particular growth regime assumed in a model of oligarchic growth and the role of gravitational focusing, that there is up to two orders of magnitude difference in the timescale of accretion at 1\,AU. 

The characteristic accretion time for the Earth is recognized as a significant unknown \citep{Kleine2009}. We calculated a series of plausible growth curves in Figure~\ref{fig:accretion} (inset) assuming the exponential growth function. With $\tau=10$\,Ma (red curve), the Earth is virtually fully ($>\!99\%$) accreted at $\sim\!50$ million years after $t_\text{CAI}$, approximately at a plausible timing for a putative {\em Giant impact} event that lead to Moon formation \citep{Barboni2017, hosono2019}. 

 
The calculated radiogenic power of the Earth is plotted as a function of accretion time (Figure~\ref{fig:accretion}). The peak radiogenic heating occurs at about 1 to 5 million years after $t_\text{CAI}$, equivalent to the time scale for Mars accretion, when the proto-Earth produces $5\times10^{3}$ to $5\times10^{4}$\,TW of power, mostly from the decay of $^{26}$Al. This power is added on top of the kinetic energy deposited by impacts during accretion.
 
Some core formation models, particularly those invoking continuous metal-silicate segregation, suggest a mean age of core separation of $\sim\!10$ million years after $t_\text{CAI}$ \citep{Kleine2009}. At this time the combined heat production from $^{26}$Al, $^{60}$Fe and $^{40}$K accounts for $\sim\!80\,\%$ (40\%, 20\% and 20\%, respectively) of the $\sim\!300$\;TW of radiogenic power in the Earth. Between 10 and 15 million years after $t_\text{CAI}$, heat production from $^{60}$Fe exceeds that of $^{26}$Al and some of the long-lived radionuclides, despite the recent low estimate for the initial ($^{60}$Fe/$^{56}$Fe)$_i$ of $(1.01\pm0.14)\times10^{-8}$ \citep{tang2015low}, a value that has been latter supported by \cite{Trappitsch2018}. Results in Figure~\ref{fig:accretion}, which assumes an instantaneously formed Earth, show that the upper bound on radiogenic energy supplied by $^{26}$Al is equivalent to 5.5$\;\times\;$10$^{31}$ J, which is comparable (within a factor of a few) to the planet's gravitational binding energy. The second most powerful short-lived radionuclide $^{60}$Fe supplies a factor of $\sim$700 times less radiogenic energy. These findings leave little doubt as to the early hot start of the Earth and the likely melting temperatures experienced by both the forming Fe-rich core and surrounding silicate mantle. Moreover, isolating $\geq\!90\%$ of the Earth's iron into the core at this time results in a super-heated condition, given contributions from radiogenic, impact, and differentiation sources.

\begin{figure}[h]
\centering
\includegraphics[width=1\textwidth]{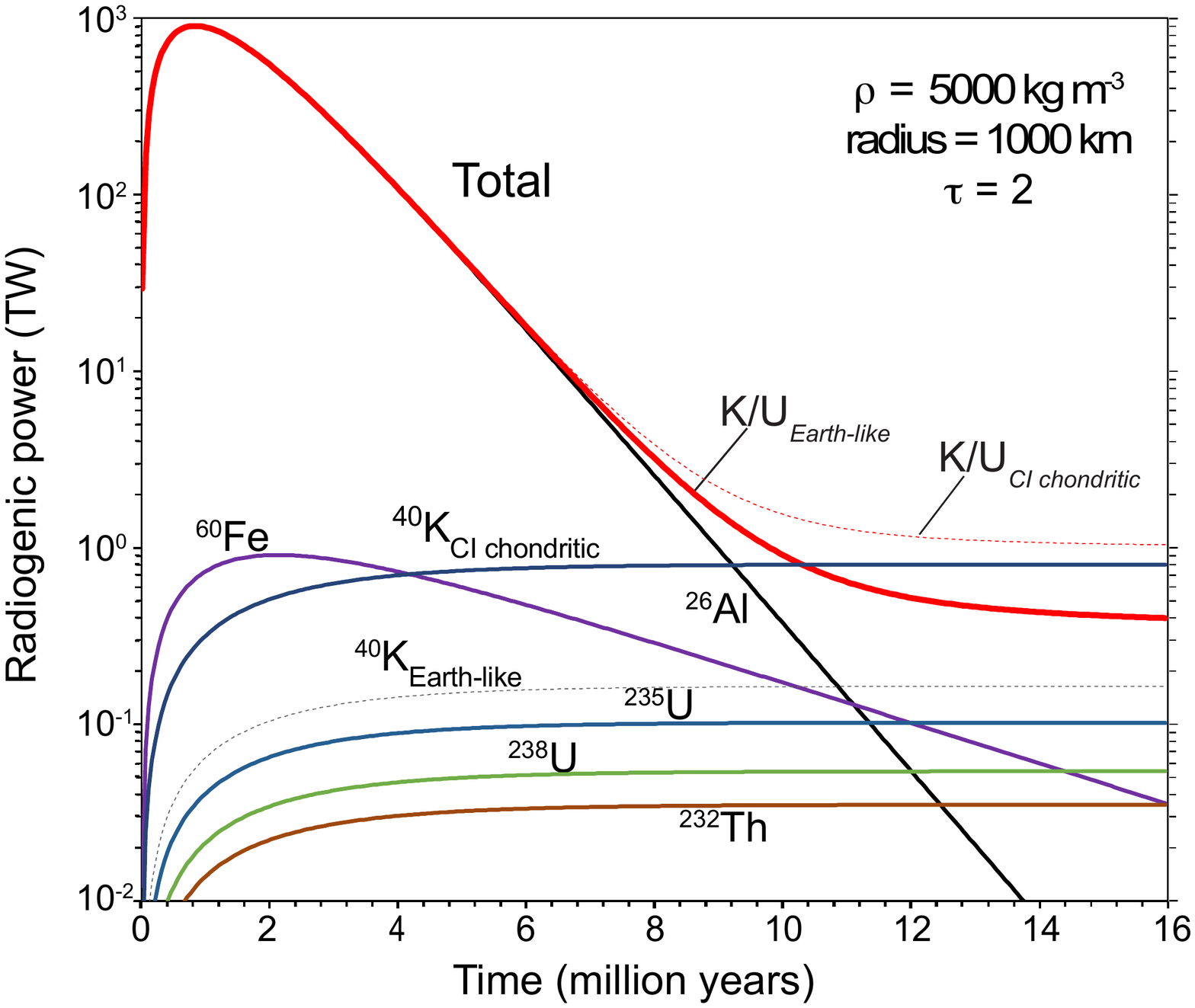}
\caption{A plot of the relative contributions of radiogenic heat to a model terrestrial body (i.e., planet or asteroid) during accretion over the first 16 million years of solar system history. Two model compositions are shown for K/U = 14,000 (Earth-like, solid lines) and K/U = 69,000 (CI chondrite, dashed lines); both models assume Fe/Al = 20 (weight ratio) and refractory elements at about 2 times that in CI chondrite, which is equivalent to an Earth-like water and CO$_2$ budget.  The terrestrial body is modeled as having a $\tau$ value of 2 and a density of 5,000 kg/m$^3$. The left y-axis (radiogenic power) is for a body with a 100 km radius. The radiogenic power scales with the body size (assuming the same composition and density); for example, for a body with 1/10 of radius, hence 1/1000 of volume, the radiogenic power will be 3 orders of magnitude smaller.} 
\label{fig:model}
\end{figure}

Figure~\ref{fig:model} presents the heat production for two different bulk compositional models of small terrestrial planets (or asteroidal body). One model assumes a bulk Earth-like composition \citep{McDonough2014} with an Fe/Al\,=\,20 and depletions in moderately volatile elements, while the other model assumes the same composition, except with moderately volatile elements set by a CI chondrite K/U value of 69,000 \citep{barrat2012geochemistry}. A small difference in heat production for these two models in the first 15 million years of solar system history is revealed. 
 
The potential of melting of a body accreted in the first 2 million years of solar system origins depends on its specific power $h$ in W/kg, where $h$ compares to $C_P \Delta T/\Delta t$, giving us the temperature increase of a body due to its radiogenic heating of $\Delta T$ over a time period of $\Delta t$. For a 100 km radius body (i.e., a size commensurate with estimates of some parent bodies of iron meteorites \citep{Goldstein2009}), a $\tau$ of 0.5 to 1 million years, and 100\,nW/kg of average specific radiogenic heating $\tilde{h}$ (note the present-day value for bulk Earth in 3\,pW/kg) would increase the body's temperature by some 3000\,K ($\Delta T$) in 1\,million years ($\Delta t$), according to a simple balance $\tilde{h}=C_P \Delta T/\Delta t$, where $C_P$ is specific heat capacity (assuming a planetary value of $\sim$1000 J$\,\text{kg}^{-1}\,\text{s}^{-1}$). This temperature increase is sufficient to induce melting and enhance the effectiveness of metal-silicate fractionation, although the actual thermal evolution also depends on the ability of the growing body to get rid of its heat, and therefore its growth curve of accretion \citep{Sramek2012}. With a 100\,km radius body, one would expect molten core and mantle \citep{Kleine2009}.

\section{Geoneutrino flux vs. radiogenic power}

There is a positive correlation between the Earth's radiogenic power and its geoneutrino flux, with the former given in TW and the latter given in number of (anti)neutrinos per cm$^2$ per second (cm$^{-2}$\,s$^{-1}$). Measurements of the geoneutrino signal can therefore place limits on the amount of Earth's radiogenic power, or Th and U abundance, and consequently constrain other refractory lithophile element concentrations \citep[e.g.,][]{McDonough2017geonu}. The Earth's geoneutrino signal is often reported in units of TNU, which stands for terrestrial neutrino units, and is the number of geoneutrinos counted over a 1-year exposure in an inverse beta decay detector having $10^{32}$ free protons ($\sim\!1$\,kiloton detector of liquid scintillation oil) and $100\%$ detection efficiency. The conversion factor between signal in TNU and flux in $\text{cm}^{-2}\,\mu\text{s}^{-1}$ depends on the Th/U ratio and has a value of $0.11\,\text{cm}^{-2}\,\mu\text{s}^{-1}\,\text{TNU}^{-1}$ for Earth's (Th/U)$_{mass}$ of 3.8.

\begin{figure}[h]
\centering
\includegraphics[width=1\textwidth]{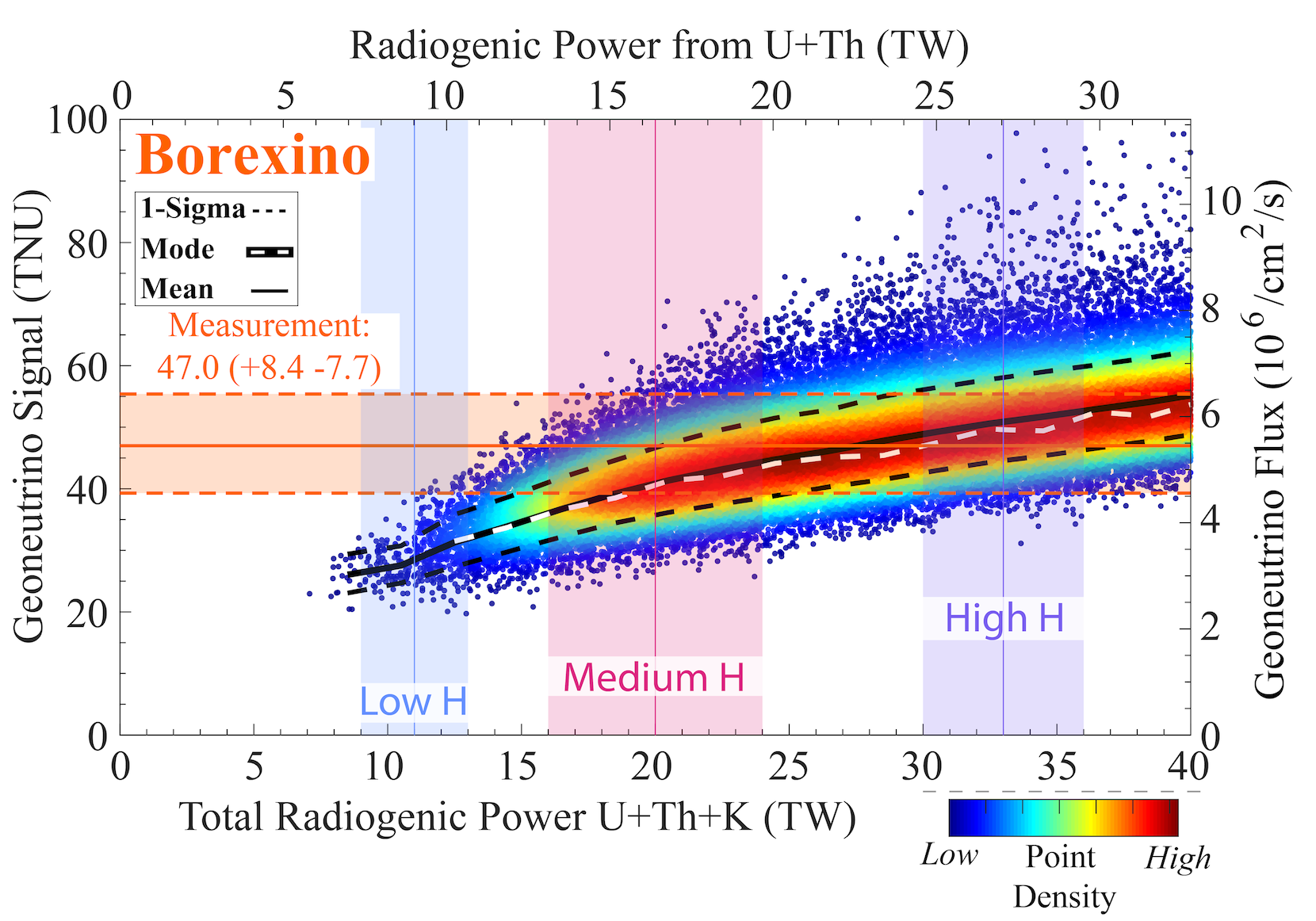}
\caption{The TNU signal (left y-axis) or geoneutrino flux (in cm$^{-2}$ s$^{-1}$; right y-axis) for the Borexino experiment versus the total radiogenic power (bottom x-axis) or only radiogenic power from Th + U (upper x-axis)(in TW) within the modeled BSE. The sloped array of points are the predictions generated with a Monte Carlo model using a reference lithosphere of the local and global contributions to the total geoneutrino flux for the Borexino location \citep{Wipperfurth2020}. The minimum solution (leftmost points) is set by the $8.1 ^{+2.7}_{-2.0}$\,TW continental lithospheric model and negligible radiogenic power in the mantle. Measured data reported by the Borexino experiment (horizontal red band) is from \cite{agostini:2020}. The low-H (blue), medium-H (pink), and high-H (purple) bands represent predictions of the BSE heat production \citep[see][for details]{Wipperfurth2020}. The 1-sigma, mean, and mode are calculated in bins every two TW.}
\label{fig:TNU}
\end{figure}

The Borexino experiment reported the geoneutrino signal \citep{agostini:2020}, based on a fixed (Th/U)$_{molar}$ = 3.9. We conducted a Monte Carlo (MC) simulation to determine the total signal at the Borexino experiment, based on a reference lithospheric model for the local and global contributions to the total flux \citep{Wipperfurth2020}. We assumed the following architecture of the BSE: lithosphere underlain by the Depleted Mantle (source of mid-ocean ridge basalt, MORB), with an underlying Enriched Mantle (source of ocean island basalts, OIB) and the volume fraction of Depleted Mantle to Enriched Mantle is 5:1 \citep{arevalo2013simplified}. Input assumption for the MC simulation include: (1) U abundance in the BSE (6 to 40 ng/g), (2) BSE (Th/U)$_{mass}$ \citep[$3.776^{+0.122}_{-0.075}$;][]{wipperfurth:2018} and (K/U)$_{mass}$ \citep[$13,800\pm1,300$;][]{Arevalo2009}, and (3) accept results with abundances of $\text{U}_\text{Depleted Mantle} \leq \text{U}_\text{Enriched Mantle}$. Figure~\ref{fig:TNU} shows the intersection of the MC model and the measured signal; the ensemble of acceptable BSE models includes the intersection of the best fit line (MC results) and the measurement field determined by the Borexino experiment (i.e., total power of 16 to 38 TW).

\cite{Wipperfurth2020} conducted a similar analysis to that above for the KamLAND signal. The KamLAND experiment recently reported their geoneutrino signal of $34.9^{+6.0}_{-5.4}$ \citep{Watanabe2016}, also based on a fixed (Th/U)$_{molar} = 3.9$. The observed ensemble of acceptable BSE models (i.e., intersection of the best fit line (MC results) and the measurement field) determined by the KamLAND experiment is between 16 and 25 TW. A combined KamLAND and Borexino result favors an Earth with $\sim$20 TW present-day total power (or a $\sim$16 TW Earth for just Th and U power; see bottom and top x-axes on Figure~\ref{fig:TNU}).

\section{Conclusions}

The Earth is a hybrid vehicle driven by primordial fuel (residual energy of accretion and core formation) and nuclear fuel (the energy given off by radioactive decay). These two sources of fuel drive the dynamical processes of convection (core and mantle), melting and volcanism. Unfortunately, we lack a gauge for either fuel source. Thus, geologists have seized the opportunity offered by particle physics to build the fuel gauge for how much nuclear power remains in the Earth. Neutrino geoscience is the emerging field of measuring and interpreting the Earth's geoneutrino emission (i.e., detecting these near massless and chargeless particles produced during radioactive beta decays). Experiments in Japan (KamLAND) and Italy (Borexino) have reported their signal and geological models have been constructed to understand the Earth's flux of geoneutrinos \citep{Wipperfurth2020}. Despite collectively having more than 25 years of data taking and more than a thousand billion billion (10$^{21}$) geoneutrinos passing through these kilometer deep detectors, these two instruments have detected less than 200 events. In the coming decade the addition of newer and larger detectors in Canada and China to the global array of geoneutrino detectors will be transformative. The annual geoneutrino count rate is predicted to increase by at least a factor of 10 \citep{sramek2016revealing}, providing the much needed statistics to interrogate the Earth. 

The Earth's geoneutrino emission scales with the amount of heat produced from radioactive decays. Surface variations in the Earth's flux of geoneutrinos \citep{Sramek2013,Usman2015} are due to the abundance and distribution of these radioactive elements in the Earth. The details of the energy deposited in the Earth and its flux of geoneutrino that removes energy from the Earth have been reviewed in detail, including an accounting of the $^{40}$K branch decay scheme, where we used an updated $\beta^-$ energy spectrum from physics and an updated branching ratio from geological studies. Whereas recent thermal models that calculate heat produced during radioactive decay agree at the level of uncertainties (i.e., $\sim\pm0.5\%$, mostly due to uncertainties in decay constants and branching ratios), models predicting the Earth's total flux of geoneutrinos differ by more than 60\%.  Using factors and equations presented here, one can calculate models for the Earth's thermal emission and (anti)neutrino luminosity, as well as that for other terrestrial bodies in the solar system and exoplanets. In the first 10 Ma of Earth's formation, the amount of radiogenic energy supplied by $^{26}$Al is equivalent to 5.5$\;\times\;$10$^{31}$ J, comparable (within a factor of a few) to the planet's gravitational binding energy. The second most powerful short-lived radionuclide $^{60}$Fe (\textit{t$_{1/2}$} = 2.6 Ma) supplies a factor of $\sim$700 times less radiogenic energy.

\acknowledgments
WFM gratefully acknowledges NSF support (EAR1650365), SAW acknowledges support from NSF (EAPSI \#1713230) and Japanese Society for the Promotion of Science (SP17054), and O\v S acknowledges Czech Science Foundation support (GA\v CR~17-01464S) for this research. We thank the many geo-, cosmo-, nuclear, and particle physics friends who have helped us to understand many of the details of these calculations, especially, B\' eda Roskovec, Steve Dye, John Learned, Sanshiro Enomoto, Hiroko Watanabe, Katherine Bermingham, and Richard Ash. Review comments by Andy Davis, Thomas Ruedas, Herbert Palme and anonymous reviewer 3 are greatly appreciated. We are grateful to Janne Blichert-Toft for her comments and editorial efforts.

\section*{Data availability}
The calculations performed in this study are openly available in the form of a Jupyter Notebook in 4TU.ResearchData repository at
\newline
\url{https://doi.org/10.4121/uuid:d635e12b-6110-44aa-8afb-6ed02070a39f}

\section*{Authors contribution}
WFM, SAW, and O\v S proposed and conceived of various portions of this study and also independently conducted heat production and luminosity calculations. O\v S calculated the $\beta$ decay spectra for the SLR and $^{40}$K. All authors contributed to the interpretation of the results. The manuscript was written by WFM, with edits and additions from O\v S, and SAW. All authors read and approved of the final manuscript.

\clearpage


\clearpage

\begin{table}[h]
\caption{Extant long-lived radioactive decay systems }
\label{tab:systems}
\setlength{\tabcolsep}{12pt}
\begin{tabular}{ccccr}
\hline
Decay system             & {Mole frac. (\%)} & Decay mode               & $\lambda$ ({a}$^{-1}$)*  & $Q$ (MeV) \\
\hline
$^{40}$K $\rightarrow\!^{40}$Ar    & 0.01167 & $\varepsilon$ (10.56\%)  & $5.810\times10^{-11}$  & 1.504 \\
$^{40}$K $\rightarrow\!^{40}$Ca    & 0.01167 & $\beta^-$ (89.44\%)      & $4.910\times10^{-10}$  & 1.311 \\
$^{40}$K overall                   &         &                          & $5.491\times10^{-10}$  & (total) 1.331 \\
$^{87}$Rb $\rightarrow\!^{87}$Sr   & 27.83   & $\beta^-$                & $1.397\times10^{-11}$  & 0.2823 \\
$^{138}$La $\rightarrow\!^{138}$Ce & 0.0888  & $\beta^-$ (34.8\%)       & $2.34\times10^{-12}$   & 1.047 \\
$^{138}$La $\rightarrow\!^{138}$Ba & 0.0888  & $\textrm{EC}$ (65.2\%)   & $4.39\times10^{-12}$   & 1.740 \\
$^{138}$La overall                 &         &                          & $6.73\times10^{-12}$   & (total) 1.504 \\
$^{147}$Sm $\rightarrow\!^{143}$Nd & 14.993  & $\alpha$                 & $6.524\times10^{-12}$  & 2.311 \\
$^{176}$Lu $\rightarrow\!^{176}$Hf & 2.598   & $\beta^-$                & $1.867\times10^{-11}$  & 1.194 \\
$^{187}$Re $\rightarrow\!^{187}$Os & 62.60   & $\beta^-$                & $1.669\times10^{-11}$  & 0.0025 \\
$^{190}$Pt $\rightarrow\!^{186}$Os & 0.0129  & $\alpha$                 & $1.415\times10^{-12}$  & 3.252 \\
$^{232}$Th $\rightarrow\!^{208}$Pb & 100     & 6$\alpha$ and 4$\beta^-$ & $4.916\times10^{-11}$  & (total) 42.646 \\
$^{235}$U $\rightarrow\!^{207}$Pb  & 0.72049 & 7$\alpha$ and 4$\beta^-$ & $9.8531\times10^{-10}$ & (total) 46.397 \\
$^{238}$U $\rightarrow\!^{206}$Pb  & 99.2740  & 8$\alpha$ and 6$\beta^-$ & $1.5513\times10^{-10}$  & (total) 51.694 \\
\hline
\end{tabular}
\par\medskip
*We follow the recommendation of  \cite{holden2011iupac} and use a (annus) for years.
Decay energy $Q$ calculated from mass differences between parent and final daughter nuclide mass data from \cite{Wang-Meng}; see Table~\ref{tab:comparison} for details on decay constant $\lambda$. $^{40}$K and $^{138}$La undergo branched decays; $Q$ entries for $^{232}$Th, $^{235}$U, and $^{238}$U account for the decay networks down to Pb nuclides. Mole fraction of U isotopes calculated from $\text{U}=\;^{238}\text{U}+\;^{235}\text{U}+\;^{234}\text{U}$;  $^{238}\text{U}/^{235}\text{U} = 137.786 \pm 0.011$ \citep{connelly2017pb}; $^{234}\text{U}/^{238}\text{U} = (5.497 \pm 0.002) \times 10^{-5}$ \citep{Villa2016}.
\end{table}

\begin{sidewaystable}
\caption{Comparison of half-life values ($t_{1/2}$, in years) of long-lived radionuclides. }
\label{tab:comparison}
\setlength{\tabcolsep}{12pt}
\begin{tabular}{c|c|ccl|c}
\hline    
 & {\textbf{NNDC}}&  {\textbf{Geochronology }} &  &  & \textbf{NNDC vs. Geo} \\
\hline
Nuclide & $t_{1/2}(\pm)$ & $t_{1/2}(\pm)$ & $\%\pm$ & Geochronology Ref. & \% rel. difference \\
\hline
$^{40}$K   & $1.248(3)\times10^9$    & $1.262(2)\times10^{9}$    & 0.2  & \cite{Naumenko2018}          & $-1.1$ \\
$^{87}$Rb  & $4.81(9)\times10^{10}$  & $4.961(16)\times10^{10}$  & 0.3  & \cite{Villa2015}             & $-3.0$ \\
$^{138}$La & $1.03(1)\times10^{11}$  & $1.03(2)\times10^{11}$    & 1.9  & \cite{Sato1981,Tanimizu2000} &  0 \\
$^{147}$Sm & $1.07(1)\times10^{11}$  & $1.063(5)\times10^{11}$    & 0.5  & \cite{Begemann2001,tavares2018toward} &  0.7 \\
$^{176}$Lu & $3.76(7)\times10^{10}$  & $3.713(16)\times10^{10}$  & 0.4  & \cite{Soderlund2004,hult2014half} &  1.3 \\
$^{187}$Re & $4.33(7)\times10^{10}$  & $4.153(8)\times10^{10}$    & 0.2  & \cite{smoliar1996re,selby2007assessment} &  4.3 \\
$^{190}$Pt & $6.5(3)\times10^{11}$   & $4.899(44)\times10^{11}$   & 0.9  &  \cite{Cook2004,Braun2017}   &  33 \\
$^{232}$Th & $1.40(1)\times10^{10}$  & $1.41(1)\times10^{10}$    & 0.7  & \cite{farley1960}            & $-0.7$ \\
$^{235}$U  & $7.038(5)\times10^{8}$  & $7.0348(20)\times10^{8}$  & 0.03 &  \cite{Villa2016}        &  0.05 \\
$^{238}$U  & $4.4683(24)\times10^{9}$ & $4.4683(96)\times10^{9}$ & 0.2  & \cite{Villa2016}             &  0 \\
\hline
\end{tabular}
\par\medskip
($\pm$) values with $t_{1/2}$ represent the absolute uncertainty in the last reported significant figure, $\%\pm$ is the relative uncertainty. The relative difference between NNDC (National Nuclear Data Center: \url{www.nndc.bnl.gov}) and geochronology values are listed in the last column.
\end{sidewaystable}

\clearpage
\vspace{-1cm}
\begin{table}[ht]
\centering
\small
\setlength\tabcolsep{2pt}
\caption{Long-lived radioactive decay systems in the Earth.}
\label{tab:hprod}
\begin{tabular}{lcccccc}
\hline
 & $^{238}$U & $^{235}$U & $^{232}$Th & $^{40}$K & $^{87}$Rb & $^{147}$Sm \\
\hline
Decay mode & $\alpha$, $\beta^-$ chain & $\alpha$, $\beta^-$ chain & $\alpha$, $\beta^-$ chain & $\beta^-$ or $\varepsilon$ & $\beta^-$ & $\alpha$ \\
Natural mole frac.$^\#$ & 0.992740 & 0.0072049 & 1.0000 & 1.167$\times$ 10$^{-4}$ & 0.2783 & 0.14993 \\
Nuclide mass (g\,mol$^{-1}$) & 238.0508  & 235.0439  & 232.0381  & 39.9640   & 86.9092   & 146.9149 \\
Standard atomic weight & & & & & & \\
\hspace{0.5cm}(Ar$_{std}$, g\,mol$^{-1}$) &  238.0289 & 238.0289 & 232.038 & 39.098 & 85.468 & 150.362 \\
Decay constant  $\lambda$ ($10^{-18}$\,s$^{-1}$) & 4.916 & 31.223 & 1.56 & 17.40 & 0.4428 & 0.2066 \\
Decay constant   $\lambda$ (a$^{-1}$) & $1.5513\times10^{-10}$ & $9.8531\times10^{-10}$ & $4.92\times10^{-11}$ & $5.491\times10^{-10}$ & $1.397\times10^{-11}$ & $6.539\times10^{-12}$ \\
Half-life $t_{1/2}$ (10$^{9}$\,a)* & 4.4683 & 0.70348 & 14.1 & 1.262 & 49.61 & 106.3 \\
$1\sigma$ uncertainty on $t_{1/2}$& & & & & & \\
\hspace{0.5cm} (10$^{9}$\,a) & 0.0096 & 0.00020 & 0.1 & 0.002 & 0.16 & 0.5 \\
$n_\alpha$ ($\alpha$ particles per decay) & 8 & 7 & 6 & 0 & 0 & 1 \\
$n_{\APnue}$ (antineutrinos per decay) & 6 & 4 & 4 & 0.8944 & 1 & 0 \\
$n_{\Pnue}$ (neutrinos per decay) & 0 & 0 & 0 & 0.1056 & 0 & 0 \\
$Q$ (MeV)$^\dagger$ & 51.694 & 46.397 & 42.646 & 1.3313 & 0.2823 & 2.3112 \\
$Q$ (pJ) & 8.2823 & 7.4335 & 6.8326 & 0.2133 & 0.0452 & 0.3703 \\
$Q_\nu$ (MeV) & 4.050 & 2.020 & 2.230 & 0.655 & 0.200 & 0 \\
$Q_\nu$ (pJ)$^\ddagger$ & 0.649 & 0.324 & 0.357 & 0.105 & 0.032 & 0 \\
$Q_h$ (MeV) & 47.6 & 44.4 & 40.4 & 0.676 & 0.082 & 2.311 \\
$Q_h$ (pJ) & 7.633 & 7.110 & 6.475 & 0.108 & 0.013 & 0.370 \\
\hline
Element mass frac. (kg/kg)** & $2.00\times10^{-8}$ & $2.00\times10^{-8}$ & $7.54\times10^{-8}$  & $2.80\times10^{-4}$  & $6.00\times10^{-7}$  & $4.06\times10^{-7}$ \\
Nuclide mass frac. (kg/kg)** & $1.99\times10^{-8}$ & $0.0144\times10^{-8}$ & $7.54\times10^{-8}$ & $3.276\times10^{-8}$ & $1.67\times10^{-7}$ & $6.09\times10^{-8}$ \\
$l'_{\APnue}$ (kg-element$^{-1}$\,s$^{-1}$) & \multicolumn{2}{c}{$7.636\times10^{7}$}  & $1.617\times10^{7}$ & $2.797\times10^{4}$ & $8.682 \times10^{5}$ & 0 \\
$L_{\APnue}$ (s$^{-1}$) & $5.99\times10^{24}$ & $1.84\times10^{23}$ & $4.93\times10^{24}$ & $3.17\times10^{25}$ & $2.11\times10^{24}$ & 0 \\
\% contribution to total $L_{\APnue}$$^\S$ & 12\% & 0.38\% & 10\% & 65\% & 4.3\% & 0 \\
$L_{\Pnue}$ (s$^{-1}$) & 0 & 0 & 0 & $3.74\times10^{24}$ & 0 & 0 \\
$h^*$ ($\mu$W/kg) nuclide     & 94.936     & 561.65   & 26.180    & 29.029    & 0.04082    & 0.3073      \\
$h'$ ($\mu$W/kg) element       & \multicolumn{2}{c}{98.293}      & 26.180    & 0.003387  & 0.01136    & 0.04607      \\
H (W) & $7.619\times10^{12}$ & $3.27\times10^{11}$ & $7.979\times10^{12}$ & $3.83\times10^{12}$ & $2.77\times10^{10}$  & $7.56\times10^{10}$ \\
\% contribution to total $H$ & 38.4\% & 1.6\% & 40.2\% & 19.3\% & 0.14\% & 0.38\% \\
\hline
\end{tabular}
\par\medskip
\begin{flushleft} 
$Q$ is the energy released per decay, $Q_{\nu}$ is the energy carried away by the electron antineutrino or neutrino per decay, $Q_h$ is the energy remaining to provide radiogenic heating per decay, ``Nuclide mass frac.'' and ``Element mass frac.'' are the abundances in silicate Earth within the reference Earth model (i.e., kg of nuclide or element per kg of rock), $l_{\APnue}$ and $l'_{\APnue}$ are the specific antineutrino luminosities of pure nuclide or element (i.e., number of $\APnue$ per kg of nuclide or element per second), $L_{\APnue}$ and $L_{\Pnue}$ are the antineutrino and neutrino luminosities of the Earth, $h^*$ and $h'$ are specific heat production rates of pure nuclide or element, $H$ is the radiogenic heat production of the Earth.
Mass of $^4$He is 4.002603254\,u and conversion of amu to MeV is 931.494. Mass of silicate Earth of $4.042\times10^{24}$\,kg is used to calculate $L_{\APnue}$, $L_{\Pnue}$, $H$.
$^\#$values from Table~\ref{tab:systems}; *values from Table~\ref{tab:comparison} Geochronology section;
**values from  \cite{McDonough1995,Arevalo2009} and Th/U ratio from \cite{wipperfurth:2018}.
$^\ddagger$Energy removed from the Earth by the $\bar{\nu}_{e}$ in the U and Th decay chains was calculated by integrating the anti-neutrino spectrum reported by S. Enomoto: \href{http://www.awa.tohoku.ac.jp/~sanshiro/research/geoneutrino/spectrum/}{www.awa.tohoku.ac.jp/$\sim$sanshiro/research/geoneutrino/spectrum}. $^\S\;$row totals to 99.3\%, with remainder from $^{176}$Lu, $^{138}$La and $^{187}$Re.
\end{flushleft}
\end{table}

\begin{sidewaystable}
\centering 
\tiny
\caption{Short-lived radioactive decay systems in the Earth.}
\begin{tabular}{cccccccll} 
\toprule
Decay system & Decay mode & Shape factor & $Q$ (MeV) & $Q_h$ (MeV) & $t_{1/2}^\ddagger$	& $(\tilde{h}/A)$ [nW/kg-elem] & Mole frac. ($\%$) parent nuclide	& Reference \\
 &  & $S(p,q)$ &  &  &  & at $t_\text{zero}$ (CAI) &  &  \\
\hline
$^{41}$Ca $\rightarrow\!^{41}$K    & EC	& -- & 0.4217 & 0 & 9.94(15) $\times$ 10$^4$ & 0 & ($^{41}$Ca/$^{40}$Ca)$_i$ = (4.6$\pm$1.0)$\times$10$^{-9}$ & \cite{Liu2017}	\\
$^{99}$Tc $\rightarrow\!^{99}$Ru   & $\beta^-$ & $0.54p^2+q^2$ (M15) &	0.2975 & 0.0957 & 2.111(12) $\times$ 10$^5$ & 46.71 & ($^{99}$Tc/$^{100}$Ru)$_i$ = $<$3.9$\times$10$^{-5}$ & \cite{wasserburg_asymptotic_1994} \\
$^{81}$Kr $\rightarrow\!^{81}$Br   & EC	& -- & 0.2809 & 0.0008 & 2.29(11) $\times$ 10$^5$ & -- & {\it not available} & -- \\
$^{126}$Sn $\rightarrow\!^{126}$Te & $\beta^-$, $\beta^-$ & $1^!$, NuDat$^!$ & 4.0502 & 2.8597 & 2.35(7) $\times$ 10$^5$ & $3.773\times10^4$ & ($^{126}$Sn/$^{124}$Sn)$_i$ = $<$3 $\times$10$^{-6}$ & \cite{Brennecka2017} \\
$^{36}$Cl $\rightarrow\!^{36}$Ar   & $\beta^-$ (98.1\%) & M15 & 0.7095 & 0.3343 & 3.01(2) $\times$ 10$^5$ & 0.9557 & ($^{36}$Cl/$^{35}$Cl)$_i$ = (1.9$\pm$0.3)$\times$10$^{-8}$ & \cite{Turner2013} \\
$^{36}$Cl $\rightarrow\!^{36}$S    & $\varepsilon$ (1.9\%) & NuDat$^!$ & $1.1421^\text{EC}$ & $1.5\times10^{-4}$ & 3.01(2) $\times$ 10$^5$	& $4.291\times10^{-4}$ & " & " \\
$^{79}$Se $\rightarrow\!^{79}$Br   & $\beta^-$ & $p^2+q^2$ [th.] & 0.1506 & 0.0559 & 3.26(28) $\times$ 10$^5$ & --	& {\it not available} & -- \\
\hline
$^{26}$Al $\rightarrow\!^{26}$Mg   & EC (18.3\%) & -- & 4.0044 & 0.3610 & $\lambda_\text{EC}/\lambda=0.1827$ & 2057 &  &  \\
$^{26}$Al $\rightarrow\!^{26}$Mg   & $\beta^+$ (81.7\%) & $p^4+\frac{10}{3}p^2q^2+q^4$ [th.] & 2.9824 & 2.7593 & $\lambda_{\beta^+}/\lambda=0.8173$ & $1.572\times10^4$ &  &  \\
$^{26}$Al $\rightarrow\!^{26}$Mg   & Overall &  &  & 3.1203 & $7.17(24)\times10^5$ & $1.777\times10^4$ & ($^{26}$Al/$^{27}$Al)$_i$ = (5.25$\pm$0.02)$\times$10$^{-5}$ & \cite{larsen2011evidence} \\
\hline
$^{10}$Be $\rightarrow\!^{10}$B    & $\beta^-$ & $p^4+\frac{10}{3}p^2q^2+q^4$ [th.] & 0.5568 & 0.2527 & 1.387(12) $\times$ 10$^6$ & $2.270\times10^4$ & ($^{10}$Be/$^{9}$Be)$_i$ = (5.3$\pm$0.5)$\times$10$^{-4}$ & \cite{Liu2010} \\
$^{93}$Zr $\rightarrow\!^{93}$Nb   & $\beta^-$ & $p^2+q^2$ [th.] & 0.0903 & 0.0456 & 1.61(5) $\times$ 10$^6$ & -- & {\it not available} & -- \\
$^{150}$Gd $\rightarrow\!^{146}$Sm & $\alpha$ & -- & 2.8077 & 2.8077 & 1.79(8) $\times$ 10$^6$ & -- & {\it not available} & -- \\
$^{135}$Cs $\rightarrow\!^{135}$Ba & $\beta^-$ & $0.10p^2+q^2$ (M15) & 0.2688 & 0.0615 & 2.3(3) $\times$ 10$^6$ & 119.4 & ($^{135}$Cs/$^{133}$Cs)$_i$ $\sim$2.8 $\times$10$^{-4}$ & \cite{Bermingham2014} \\
$^{60}$Fe $\rightarrow\!^{60}$Ni   & $\beta^-$, $\beta^-$ & $1^!$, NuDat$^!$ & 3.0598 & 2.7077 & 2.62(4) $\times$ 10$^6$ & 0.3635 & ($^{60}$Fe/$^{56}$Fe)$_i$ = (1.01$\pm$0.14)$\times$10$^{-8}$ & \cite{tang2015low} \\
$^{154}$Dy $\rightarrow\!^{150}$Gd & $\alpha$ & -- & 2.9451 & 2.9451 & 3.0(15) $\times$ 10$^6$ & -- & {\it not available} & -- \\
$^{53}$Mn $\rightarrow\!^{53}$Cr   & EC & -- & 0.5968 & 0 & 3.74(4) $\times$ 10$^6$ & 0 & ($^{53}$Mn/$^{55}$Mn)$_i$ = (6.28$\pm$0.33)$\times$10$^{-6}$ & \cite{Trinquier2008} \\
$^{98}$Tc $\rightarrow\!^{98}$Ru   & $\beta^-$ & $1^!$ & 1.794 & 1.5165 & 4.2(3) $\times$ 10$^6$ & 212.4 & ($^{98}$Tc/$^{96}$Ru)$_i$ $\sim$2$\times$10$^{-5}$ & \cite{becker2003search} \\
$^{97}$Tc $\rightarrow\!^{97}$Mo   & EC & -- & 0.3247 & 0 & 4.21(16) $\times$ 10$^6$ & 0 & ($^{97}$Tc/$^{98}$Ru)$_i$ = $<$4$\times$10$^{-4}$ & \cite{Dauphas2003} \\
$^{107}$Pd $\rightarrow\!^{107}$Ag & $\beta^-$ & $p^2+q^2$ [th.] & 0.0341 & 0.0133 & 6.5(3) $\times$ 10$^6$ & 0.2768 & ($^{107}$Pd/$^{108}$Pd)$_i$ = (3.5$\pm$0.05)$\times$10$^{-5}$ & \cite{matthes2018pd} \\
$^{182}$Hf $\rightarrow\!^{182}$W  & $\beta^-$, $\beta^-$ & $p^2+q^2$ [th.], NuDat$^!$ & 2.1958 & 1.8276 & 8.90(9) $\times$ 10$^6$ & 87.07 & ($^{182}$Hf/$^{180}$Hf)$_i$ = (1.018$\pm$0.022)$\times$10$^{-4}$ & \cite{Kruijer2014} \\
$^{129}$I $\rightarrow\!^{129}$Xe  & $\beta^-$ & $p^2+3.16q^2$ (M15) & 0.1889 & 0.0853 & 1.57(4) $\times$ 10$^7$ & 12.71 & ($^{129}$I/$^{127}$I)$_i$ $\sim$1.4$\times$10$^{-4}$ & \cite{Gilmour2017} \\
$^{205}$Pb $\rightarrow\!^{205}$Tl & EC & -- & 0.05067 & 0 & 1.73(7) $\times$ 10$^7$ & 0 & ($^{205}$Pb/$^{204}$Pb)$_i$ = (1.0$\pm$0.2)$\times$10$^{-3}$ & \cite{Baker2010} \\
$^{92}$Nb $\rightarrow\!^{92}$Zr   & EC & -- & 2.0059 & 1.4956 & 3.47(24) $\times$ 10$^7$ & 16.71 & ($^{92}$Nb/$^{93}$Nb)$_i$ = (1.7$\pm$0.3)$\times$10$^{-5}$ & \cite{Iizuka2016} \\
$^{244}$Pu $\rightarrow\!^{232}$Th & $3\alpha$, $2\beta^{-}$ $^\dagger$ & NuDat$^!$ & 17.0836 & 15.6264 & 8.11(3) $\times$ 10$^7$ & $1.363\times10^4$ & ($^{244}$Pu/$^{238}$U)$_i$ $\sim$0.0079 & \cite{Turner2007} \\
$^{146}$Sm $\rightarrow\!^{142}$Nd & $\alpha$ & -- & 2.5288 & 2.5288 & 1.03(5) $\times$ 10$^8$ & 87.96 & ($^{146}$Sm/$^{144}$Sm)$_i$ = (8.28$\pm$0.22)$\times$10$^{-3}$ & \cite{Marks2014,Meissner1987} \\
\hline
\end{tabular}
\label{tab:ShortLived}
\begin{flushleft} 
$Q$ is the energy of transition ($Q$ value) not accounting for possible branching; $Q_h$ is the energy that remains in the Earth to provide radiogenic heating per decay, accounting for branching. 
M15 = shape factor from \cite{Mougeot2015}.
$^!$In some cases, we use shape factors equal to 1 or NuDat-tabulated mean electron energies for forbidden transitions, due to lack of better inputs. 
$^\text{EC}$Reports the $Q$ value of EC branch. 
$^\ddagger$\,Half-lives are from NNDC (\href{http//www.nndc.bnl.gov}{www.nndc.bnl.gov}). 
Heating coefficients $(\tilde{h}/A)$ (in nW\,kg-elem$^{-1}$), so that radiogenic power per unit mass of rock $\tilde h$ can be calculated from $\tilde h = (\tilde{h}/A) \times A$, $A$ being the elemental mass fraction (kg-element/kg-rock), are obtained similarly to equation \eqref{specheat}. [Note, we report here $(\tilde{h}/A)$ in nW (c.f., \eqref{specheat}, which is given in $\mu$W) for these SLR.]
$^\dagger\,^{244}$Pu also undergoes spontaneous fission to $^{130-136}$Xe isotopes with a fission branching probability of 0.12\%.
The individual decay energies for the double $\beta^-$ steps are as follows: $^{126}\text{Sn}\rightarrow\!^{126}\text{Te}$: $Q=0.3782+3.6720=4.0502$\,MeV; 
$^{60}\text{Fe}\rightarrow\!^{60}\text{Ni}$: $Q=0.237+2.8228=3.0598$\,MeV; 
$^{182}\text{Hf}\rightarrow\!^{182}\text{W}$: $Q=0.3813+1.8145=2.1958$\,MeV. 
The individual decay energies for $^{244}\text{Pu}\rightarrow\!^{232}\text{Th}$: $Q=4.6655+0.3991+2.1901+5.2558+4.5731=17.0836$\,MeV.   
Uncertainties cited in the initial ratios of the short-lived nuclides are 1 standard deviation. 

\end{flushleft}
\end{sidewaystable}

\begin{table}
\caption{Comparison of Earth models for their total radiogenic power (H) and geoneutrino luminosity ($L_{\APnue}+L_{\Pnue}$) }
\label{tab:result}
\setlength{\tabcolsep}{12pt}
\begin{tabular}{lccccccc}
\hline

Conc. \textit{BSE}$^\dagger$ (mg/kg) & 0.020 & 0.0754 & 280 & 0.60 & 0.406 &   &  \\ \hline
H (TW) & U & Th & K & Rb & Sm & Total & $\Delta^\ddagger$ \\ \hline
\cite{Enomoto2006} & 7.97 & 8.02 & 3.73 & 0.028 & 0.076 & 19.8 & 0.2\% \\
\cite{Dye2012} & 7.97 & 8.01 & 3.77 &  -- & --  & 19.7 & 0.6\% \\
\cite{Ruedas2017} & 8.01 & 8.04 & 3.88 &  -- & --  & 19.9 & -0.3\% \\ 
this study & 7.95 & 7.98 & 3.84 & 0.028 & 0.076 & 19.9 & 0.0\% \\ \hline
($L_{\APnue}$ + $L_{\Pnue}$) $\times10^{24}$ s$^{-1}$ & U & Th & $K_{\APnue}$ & $K_{\Pnue}$ & Rb & Total & $\Delta_{\APnue}$/$\Delta_T$ \\ \hline
\cite{Enomoto2006} & 6.17 & 4.94 & 3.05 & -- &  2.2 & 16.3 & 64\%/66\% \\
\cite{Dye2012} & 6.17 & 4.94 & 30.5 &  -- &  -- & 41.6 & 7\%/14\% \\
\cite{Usman2015} & $\,\;$5.46* & 3.30 & 25.0 &  -- &  -- & 33.8 & 25\%/30\% \\ 
this study & 6.13 & 4.93 & 31.7 & 3.7 &  2.1 & 48.6 & 0\% \\ \hline
\end{tabular}

\begin{flushleft} 
\hspace{0.5cm} \textit{BSE}$^\dagger$ (bulk silicate Earth) see Table~\ref{tab:hprod} for inputs used in this study and units. \\
\hspace{0.5cm} $\Delta^\ddagger$ difference calculated relative to this study.\\ \hspace{0.5cm} $\Delta_{\APnue}$ refers to $\APnue$ from U, Th, K and Rb and $\Delta_T$ to total (anti)neutrino luminosity.\\
\hspace{0.5cm} *assumes a $1.64\times10^{23}$ $L_{\APnue}$ s$^{-1}$ contribution from $^{235}$U. 
\end{flushleft}

\end{table}

\end{document}